\documentstyle[12pt]{article}
\def\dspace{\baselineskip = .30in}
\def\strut{\rule[-.5cm]{0cm}{1cm}}
\begin{document}

\title{Supersymmetric Unification Without Proton
Decay\thanks{Supported in part by NATO Grant \#CRG-910573 and Department
of Energy Grant \#DE-FG02-91ER406267.}}

\author{{\bf G. Lazarides and C. Panagiotakopoulos}\\ Physics Division\\
School of Technology\\ University of Thessaloniki\\ Thessaloniki,
Greece\\\\ {\bf Q. Shafi}\\ Bartol Research Institute\\ University of
Delaware\\ Newark, Delaware 19716 U.S.A.}

\date { }
\maketitle

\dspace
\centerline{\bf Abstract}
\vspace{.2in}

We present a supersymmetric grand unification scheme based on the
gauge group
$SU(3)_c\times SU(3)_L\times SU(3)_R$ in which the proton and
the lightest supersymmetric particle are stable,
neutrinos are necessarily
massive, and the observed baryon asymmetry originates in the lepton
sector. Such models are also consistent with the measured value of
$\sin^2\theta_W$ as well as unification of the gauge couplings.
\newpage

The minimal supersymmetric $SU(5)$ model$^{(1)}$ predicts a value for
$\sin^2\theta_W$ which is remarkably consistent with the measured
numbers$^{(2)}$. Despite this success, however, it appears that the minimal
scheme faces an uphill task. For instance, the predicted asymptotic
mass relations involving the first two families $(m_d = m_e$ and $m_s
= m_\mu)$ are in glaring contradiction with the observations.
Furthermore, there is a close relationship between the light fermion
masses and proton decay (into $K^+\bar{\nu}_i$, etc.) mediated
through dimension five interactions. The experimental situation
regarding the latter is getting to a point$^{(3)}$ where it appears puzzling
why the spectacular events have not been observed. Additional
motivation for thinking beyond the minimal $SU(5)$ framework comes
from considerations of neutrino masses, baryon asymmetry, etc.

One modest extension would be an enlargement of the Higgs sector as
well as the introduction of singlet chiral superfields. Another
possibility is to embed in an $SO(10)$ framework.
In this paper we follow a relatively less treaded path,
in which the standard $SU(3)_c\times SU(2)_L\times U(1)_Y$ model is
embedded in $G \equiv SU(3)_c\times SU(3)_L\times SU(3)_R$. There are
additional reasons besides those already listed for studying models
based on $G$. The simplest compactification schemes$^{(4)}$ involving the
heterotic $E_8\times E_8$ superstring theory give rise to $G$,
although the present study will be largely motivated within the
ordinary GUT framework. The imposition of a $Z_2$
symmetry ensures the (almost complete) absence of baryon number
violation. Indeed, the main source of baryon number violation is the
non-perturbative sector of the standard electroweak model!
Other `advantages' include the absence of wrong fermion mass relations,
`naturally' massive neutrinos, and an elegant mechanism for
generating the observed baryon asymmetry. We will present here two
different realizations of the $G \equiv SU(3)_c\times SU(3)_L\times
SU(3)_R$ model. In the first one the symmetry breaking of $G$
to the minimal supersymmetric standard model (MSSM) proceeds in two
steps, whereas in the second scenario there is no intermediate step
breaking.

The left handed lepton, quark and antiquark superfields transform
under G as follows:

\begin{equation}\begin{array}{lcccc}
\lambda & = & (1,\bar{3},3) & = & \left( \begin{array}{ccc}H^{(1)} &
H^{(2)} &
L\\ E^c & \nu^c & N\end{array}\right),\strut\\

Q & = & (3,3,1) & = & \left( \begin{array}{c}q\\
g\end{array}\right),\strut\\

Q^c & = & (\bar{3}, 1, \bar{3}) & = & \left( \begin{array}{ccc} u^c &
d^c & g^c\end{array}\right).\end{array}
\end{equation}

\noindent
Here $N$ and $\nu^c$ denote standard model singlet superfields, while
$g$
$(g^c)$ is an additional down-type quark (antiquark). We will be working with
 five
$\lambda$'s, two $\bar{\lambda}$'s, five $Q$'s, two $\bar{Q}$'s, five $
Q^c$'s, two $\bar{Q}^c$'s and two $G$-singlet superfields $S_1, S_2$.
Notice that the field content is chosen to ensure unification above
the GUT scale.

We choose to impose invariance under the discrete $Z_2$ symmetry$^{(5)}$
$P_1$ generated by $Q_a \rightarrow - Q_a,\; Q_a^c \rightarrow - Q_a^c,\;
 \lambda_a
\rightarrow \lambda_a (a = 1,2,...,5),\;
\bar{Q}_\alpha\;\rightarrow -
\bar{Q}_\alpha, \bar{Q}^c_\alpha \rightarrow - \bar{Q}^c_\alpha,\;
\bar{\lambda}_\alpha \rightarrow \bar{\lambda}_\alpha (\alpha = 1,2)$.
This has the effect of forbidding baryon number violating couplings
$Q_{a_{1}} Q_{a_{2}} Q_{a_{3}},\; Q^c_{a_{1}} Q^c_{a_{2}}
Q^c_{a_{3}},\; \bar{Q}_{\alpha_{1}} \bar{Q}_{\alpha_{2}}
\bar{Q}_{\alpha_{3}}$ and $\bar{Q}^c_{\alpha_{1}}
\bar{Q}^c_{\alpha_{2}} \bar{Q}^c_{\alpha_{3}}$. Consequently,
proton decay mediated by
dimension four,  five and six operators cannot occur. Moreover, the
structure of $G$ is such that there is no gauge boson
mediated decay. The proton is stable.

The symmetry breaking to MSSM is obtained through superpotential
couplings of the form $S_1(\lambda_4\bar{\lambda}_1 - M^2_X)$ and
$S_2(\lambda_5\bar{\lambda}_2 - M^2_{B-L})$. The superlarge vacuum
expectation values are acquired by the following fields:

\begin{equation}\begin{array}{ccccc}
<N_4> & = & <\bar{N}_1>^* & = & M_X,\\
<\nu^c_5> & = & <\bar{\nu}^c_2>^* & = & M_{B-L}.\end{array}
\end{equation}

We also introduce in the superpotential explicit mass terms of order
$M_X$ for $Q,\bar{Q}$ and $Q^c,\bar{Q}^c$ pairs.
In the first realization of the
theory, we assume that $M_{B-L} \ll M_X$. This means that the breaking
to MSSM proceeds in two steps:

\begin{equation}
G \stackrel{_{\longrightarrow}}{_{M_X}} SU(3)_c\times
SU(2)_L\times SU(2)_R\times U(1)_{B-L}
\stackrel{_{\longrightarrow}}{_{M_{B-L}}} MSSM \; .
\end{equation}

\noindent
We will shortly provide estimates for the scales $M_X$ and $M_{B-L}$.

In addition to the $Z_2$ symmetry $P_1$, we impose an additional $Z_2$
symmetry $P_2$, under which $\lambda_5\rightarrow -\lambda_5$,
$\bar{\lambda}_2 \rightarrow - \bar{\lambda}_2$, while all other fields
remain  invariant. This symmetry, combined with the $Z_2$ subgroup
generated by the element (diag $(-1, -1)$, diag $(-1,-1))$ of
$SU(2)_L \times SU(2)_R$, remains unbroken by all the vacuum expectation
values and acts as `matter parity'. The purpose of introducing this
symmetry is to eliminate some lepton number violating couplings at the
level of the MSSM. This is necessary for  the implemention of our
mechanism of generating the baryon asymmetry of  the Universe. The
physical consequences of matter parity will become clear in the
following.

We next study the spectrum of the theory for the case
$M_{B-L} \ll M_{X}$.
The mass matrices will be
classified according to the quantum numbers of the states under the
gauge group $SU(3)_c \times SU(2)_L \times U(1)_Y$ and  the $Z_2$
matter parity. We will also confine  ourselves to the discussion of
the fermionic mass matrices. The result applies to the bosonic
components as well, modulo corrections of the order of the
supersymmetry breaking scale $M_S \sim 1 TeV$.

Let us first look at the mass matrix of the colorless $SU(2)_L$ -
doublet fields with positive matter parity. The  bosonic partners of
these fields have the correct quantum numbers to play the role of the
electroweak Higgs doublets. The mass matrix couples $H^{(2)}_{m}$ $(m =
1,2,3,4)$, $\bar{H}^{(1)}_{1}$ and $L_5$ with $H^{(1)}_{m} (m = 1,2,3,4)$,
$\bar{H}^{(2)e}_{1}$ and $\bar{L}_2$. In the limit $M_{B-L} = 0$, $L_5$ and
$\bar{L}_2$ are exactly massless and
decouple from the rest of the mass matrix.
The remaining mass
matrix is symmetric with all eigenvalues typically $\sim M_X$.
Mixings between $H^{(2)}_{m}$ and $\bar{H}^{(1)}_{1}$, as well as between
$H^{(1)}_{m}$ and $\bar{H}^{(2)}_{1}$, are typically $\sim M_X/M_P$
($M_P \simeq 1.2 \cdot 10^{19} GeV$ is the Planck mass),
whereas mixings among the
$H^{(2)}_{m}$'s and among the $H^{(1)}_{m}$'s are of order unity.
Diagonalization of this symmetric mass matrix by a unitary
transformation gives a positive matrix, diag $(M_1, M_2, M_3, M_4,
\bar{M}_1)$, which predominantly couples four linear combinations of the
four $H^{(2)}_{m}$'s with the corresponding four linear combinations of
the four $H^{(1)}_{m}$'s as well as $\bar{H}^{(1)}_{1}$ with
$\bar{H}^{(2)}_{1}$. To obtain a unique light pair of electroweak Higgs
doublets,  $h^{(2)}$, $h^{(1)}$, we must fine  tune one eigenvalue among the
four $M's$ so that it becomes $ \sim M_S$.
Thus we reach the important conclusion that, neglecting mass
contributions which vanish in the $M_{B-L}=0$ limit,
the pair of electroweak doublets, $h^{(2)},
h^{(1)}$, predominantly lies in one linear combination $\lambda_0$ of
$\lambda's$. Removing the superheavy states with masses of order
$M_X$, we are left with a $ 2 \times 2 $ mass matrix which
couples $h^{(2)}$ and $L_5$ with
$h^{(1)}$ and $\bar{L}_2$.  As $M_{B-L}$ grows, the $ L_5 -h^{(1)},
h^{(2)} - \bar{L}_2$ and $L_5 - \bar{L}_2$ mass terms become of the order
of $ M_{B-L}, M_X M_{B-L} /M_P$ and $M^2_{B-L}/ M_P$
respectively. In order to keep the pair of electroweak doublets,
$h^{(2)}, h^{(1)}$, essentially unaffected by these contributions,
we must then fine tune the off-diagonal mass terms so that both of them
are much smaller than $\sim M^2_{B-L}/M_P$ and their product is lower
than $\sim M_S M^2_{B-L}/M_P$.

The mass matrix of the colorless $SU(2)_L$ - doublets with negative
matter parity couples $H^{(2)}_{5}$, $\bar{H}^{(1)}_{2}$ and $L_m$ $(m
= 1,2,3,4)$ with $H^{(2)}_{5}$, $\bar{H}^{(2)}_{2}$ and $\bar{L}_{1}$.
The fields $H^{(2)}_{5}$ and $\bar{H}^{(1)}_{2}$ predominantly pair up with
$H^{(1)}_{5}$ and $\bar{H}^{(2)}_{2}$ to form states with
masses $\sim M_X.$ Also, one combination  of the four $L_m$ 's pairs
up with $\bar{L}_{1}$ to form a state with mass $\sim
M^{2}_{X}/M_{P}$. The remaining three orthogonal linear combinations $l_{i} (i
=
1,2,3)$ of the four $L_m$ 's remain massless and play the role of
the ordinary light lepton doublets.

The $SU(2)_L$ - singlet leptons with positive matter parity are
$E^{c}_{5}$ and $\bar{E}^{c}_{2}$. They form a state with
mass $\sim M^2_{B-L}/M_P$. One linear combination of the four
negative matter parity singlet  leptons $E^{c}_{m}$ $(m = 1,2,3,4)$
pairs up with $\bar{E}^{c}_{1}$ to form a state with mass
$\sim M^{2}_{X}/ M_P$. The three remaining orthogonal linear combinations
$e^{c}_{i}$ $(i = 1,2,3)$ which belong, to a very good approximation,
to the same $\lambda 's$ as the light $l_i$ $(i = 1,2,3)$, remain
massless and play the role of the ordinary light $SU(2)_{L}$ -singlet
leptons.

In the positive matter parity neutral sector $(N_{m} (m = 1,2,3,4),
\bar{N}_{1}, \nu^{c}_{5}, \bar{\nu}^{c}_{2})$, all states get masses
$\sim M^{2}_{X} /M_P$ except for the two linear combinations acquiring
the large expectation values  which get masses of the order of the
corresponding expectation value ($M_X$ and $M_{B-L}$, respectively) and
one combination of $\nu^c_5$ and $\bar{\nu}^c_2$ which gets a mass
$\sim M^2_{B-L}/M_P$.
In the
negative matter parity neutral sector $(N_{5}, \bar{N}_{2}, \nu^{c}_{m}
(m = 1,2,3,4), \bar{\nu}^{c}_{1})$, $N_{5}, \bar{N}_{2}$ get masses
$\sim M^{2}_{X}/M_{P}$ and have mixings $\sim M_{B-L}/M_{X}$ with
$\nu^{c}_{m}$ 's and $\bar{\nu}^{c}_{1}$. One linear combination
of the four $\nu^{c}_{m}$ 's (let us say $\nu^{c}_{4}$) pairs up with
$\bar{\nu}^{c}_{1}$ to form a state with mass $\sim M^{2}_{X}/M_P$. The
remaining three linear combinations $\nu^{c}_{i} (i = 1,2,3)$, which
are $SU(2)_R$ - partners of the light $e^{c}_{i}$ 's acquire masses $\sim
M^{2}_{B-L}/M_P$. The mass matrix of the light $\nu^{c}_{i}$ 's is
symmetric and therefore can be diagonalized by a single unitary
transformation. In the following we will assume that the mass matrix
of the $\nu^{c}_{i}$ 's is  already diagonal in the form diag
$(M_{{\nu}^{c}_{1}}, M_{{\nu}^{c}_{2}}, M_{{\nu}^{c}_{3}})$.

The quark sector is simple. After pairing two $Q,\bar{Q}$ and two
$Q^c, \bar{Q}^c$ pairs with mass terms assumed of order $10^{-2}$ $M_X$, we
are left with three massless doublets $q_i = \left( \begin{array}{l}u_i\\
d_i\end{array}\right)$, as well as three massless $u_i^c$'s
and three massless $d_i^c$'s belonging to the same three $Q_i^c$'s
$(i = 1,2,3)$, all with negative matter parity. We also have the corresponding
 three pairs of $g_i,
g_i^c$'s $(i = 1,2,3)$ which have positive matter parity and acquire
`superheavy' masses $ \sim M_X$.

Between the energy scales $M_W$ and $M_X$, the mass spectrum is not
symmetric in the three $SU(3)$'s (see above). However, above $M_X$
the symmetry is restored and the three gauge couplings
are equal (here we assume that $G$ emerges at $M_P$ with a unique
gauge coupling constant). Using the standard one loop renormalization group
evolution of the gauge couplings, and assuming that the supersymmetry
breaking scale $M_S \simeq 10^{3} GeV$, one can obtain the following
(consistent) solution:

\begin{equation}\begin{array}{rcl}
\sin^2\theta_W (M_Z) & \approx & 0.23,\; \alpha_s (M_Z) \simeq 0.115,\;
\alpha_G(M_P) \simeq  0.158,\strut\strut\\
M_X & \approx & 10^{16.5} GeV,\; M_{B-L} \simeq 10^{13.5}GeV,\end{array}
\end{equation}

\noindent
where $\alpha_G$ denotes the unified gauge coupling above $M_X$.

The fact that the pair of electroweak doublets $h^{(2)}, h^{(1)}$
predominantly belongs to the same linear combination of $\lambda_a$'s
(namely $\lambda_0$) together with the observation that the `light'
$u_i^c$'s, $d_i^c$'s belong to the same three $Q_i^c$'s $(i = 1,2,3)$
implies that the tree-level up- and down-quark mass matrices are
proportional (they both come from the same couplings $Q_i Q_j^c
\lambda_0$). Assuming that the third generation acquires tree-level
masses, we obtain the asymptotic relation

\begin{equation}
\frac{m_t}{m_b} \simeq \tan\beta \equiv \frac{<h^{(1)}>}{<h^{(2)}>}.
\end{equation}

\noindent
To account for the correct masses and mixings of the two lighter quark
families, we must invoke some other mechanism (radiative corrections,
...) which we will not specify here. In the leptonic sector, we have
seen that the three `light' $\nu_i^c$'s and the light $e_i^c$'s
belong to the same $\lambda_i$'s $(i = 1,2,3)$. Thus, the tree-level
Dirac mass matrix of the neutrinos is proportional to the tree-level
charged lepton mass matrix (since both these mass matrices come from
the same couplings $\lambda_i\lambda_j\lambda_0)$. Assuming again
that only the third family acquires tree-level masses we obtain the
asymptotic relation

\begin{equation}
\frac{m^D_{\nu_{\tau}}}{m_\tau} \simeq \tan\beta
\end{equation}

\noindent
where the superscript $D$ denotes a Dirac neutrino mass.

So far we have seen that the model is consistent with the low
energy phenomenology. The proton is stable and $\sin^2\theta_W$ is in
the right ball park. Next we address the issue of baryon asymmetry.
Since the conventional baryon number violating couplings are missing,
we must resort to the lepton sector which contains the
lepton number violating Majorana mass terms.
The idea then is to generate an initial lepton asymmetry
$n_L/s$ through the decays of $\nu^c_i (i = 1,2,3)$ $^{(6,7)}$. The
non-perturbative sphaleronlike effects in the electroweak
sector$^{(8)}$
would then convert a fraction of the asymmetry $n_L/s$ into the
observed baryon asymmetry $n_b/s$, provided the `reheat' temperature
$T_r$ after the completion of inflation is greater than or equal to
about 100 GeV. In order to ensure the survival of
$n_L/s$, we must ensure that all lepton number violating 2-2
scatterings are out of equilibrium at temperatures between $T_r$ and
100 GeV. This is elegantly achieved with the help of the $Z_2$ matter parity.

Within an inflationary framework driven by a gauge singlet field
$\phi$ (note that inflation is needed here to eliminate the monopole
problem), we assume copious production of $\nu^c_i$ from the out of
equilibrium decay of $\phi$ as it oscillates about the minimum of the
potential. If $m_\phi \stackrel{_>}{_\sim} 2 M_{\nu^{c}_{i}} (i = 1,2,3)$,
the inflaton will predominantly decay to the
heaviest $\nu^c_i$ which we assume is $\nu_3^c$ (recall that we are in a
basis in which the $\nu_i^c$-mass matrix is diagonal). The decay width
of $\phi$ is given by

\begin{equation}
\Gamma_\phi \sim \frac{1}{16\pi} \left(
\frac{M_{\nu_{3}^{c}}}{<\phi>}\right)^2 m_\phi\;,
\end{equation}

\noindent
where $<\phi> \sim 10^{17} - 10^{18}$ GeV as estimated from the
requirement that $\delta \rho/\rho \sim 10^{-5}$. The corresponding
`reheat' temperature is given by

\begin{equation}\begin{array}{lcl}
T_r & \sim & \frac{1}{3} (\Gamma_\phi M_P)^\frac{1}{2}\strut\strut\\
& \sim & 2.6 \times 10^3 \left(\frac{M_{\nu_{3}^c}}{10^8 GeV} \right)
\left( \frac{10^{17.5} GeV}{<\phi>} \right)
\left( \frac{m_\phi}{10^{9.5}GeV }\right)^\frac{1}{2} GeV.\end{array}
\end{equation}

\noindent
We will choose $M_{\nu_{3}^{c}} \simeq  10^8$  GeV,$m_\phi \simeq
10^{9.5}$ GeV  and $<\phi> \simeq 10^{17\cdot 5}$ GeV. This implies $T_r \simeq
2.6\ TeV$, which is
higher than the minimum allowed `reheat' temperature.

A simple double-cut diagram which leads to a lepton asymmetry
through $\nu^c_i$ decays (in our case $i = 3$) is shown in Fig.1
(see also ref. 6). A straightforward
estimate leads to

\begin{equation}
n_{L_{i}}/s \sim \frac{1}{4\pi} \frac{T_r}{m_\phi}
\frac{M_{\nu_{i}^c}}{\left( \frac{m_Dm_D^+}{\mid <h^{(1)}>\mid ^2} \right)
_{ii}} Im \sum_k \left( \frac{m_Dm_D^+}{\mid <h^{(1)}>\mid ^2} \right)^2_{ik}
\frac{M_{\nu_{k}^c}}{\vec{p}^2 + M_{\nu^c_{k}}^2} \;,
\end{equation}

\noindent
where $m_D$ is the Dirac mass matrix of the neutrinos and the internal
three-momentum $|\vec{p}| \simeq \frac{1}{2} M_{\nu_{i}^{c}}$. To
help maximize the lepton asymmetry we set $M_{\nu_{2}^c} \simeq 2\cdot
10^5\ GeV$, and assume that the dominant matrix element of $m_D$ is
$(m_D)_{23}$ (the element which couples $\nu_2^c$ and $\nu_3$). Eq. (9)
then gives

\begin{equation}
n_{L_{3}}/s \sim \frac{1}{\pi} \cdot \frac{T_r}{m_\phi} \cdot
\frac{M_{\nu_{2}^c}}{M_{\nu_{3}^c}} \cdot
\left|\frac{(m_D)_{23}}{<h^{(1)}>}\right|^2.
\end{equation}

\noindent
Proportionality of $m_D$ and the charged lepton mass matrix at
tree-level implies

\begin{equation}
\frac{(m_D)_{23}}{<h^{(1)}>} \simeq \frac{m_\tau}{<h^{(2)}>} \simeq
\frac{m_\tau\tan\beta}{<h^{(1)}>} \;.
\end{equation}

\noindent
Taking $\tan\beta \simeq 44$ and $<h^{(1)}> \simeq 170\
GeV$, for example, one finds that $(m_D)_{23} / <h^{(1)}> \simeq 0.46$ and
$n_{L_{3}}/s \sim 10^{-10} \sim n_b/s$.

With this choice of the parameters, the mass of the tau neutrino
turns out to be

\begin{equation}
m_{\nu_{\tau}} \simeq \frac{(m_\tau\tan\beta)^2}{M_{\nu_{2}^c}} \sim
30 MeV \;.
\end{equation}

\noindent
The requirement of a sizeable baryon asymmetry has led us
to a $\tau$-neutrino mass in the MeV range. The lifetime and
branching ratio into radiative modes of such a neutrino is strongly
constrained by astrophysical consider- ations$^{(9)}$. A neutrino with
mass $\sim 30 MeV$ could be astrophysically acceptable if it decays with
a lifetime $\tau_{\nu_{\tau}} \leq 20 sec$ into  $e^+ e^- \nu$ or $\nu
\nu^{\prime} \bar{\nu}^{\prime \prime}$, with $\nu, \nu^\prime$,
$\nu^{\prime \prime}$ any of the two lighter neutrinos. This can
easily be achieved provided $M_S$ is somewhat
lower than $1 TeV$. The relevant diagrams are simple one loop box
diagrams involving supersymmetric partners of charged leptons or of
neutrinos and  two winos.

As another possibility,
one can consider the mode $\nu_\tau
\rightarrow \nu_e a$, where $a$ is an axion field. The decay
rate is then $\Gamma \sim \frac{1}{16\pi}
(\frac{m_{\nu_{\tau}}}{f_a})^2 m_{\nu_{\tau}}$, where $f_a$ is the
axion decay constant. For $f_a \simeq 3\cdot 10^9$GeV, we obtain
$\tau_{\nu_{\tau}} \simeq 20$ sec which is acceptable. Note that an
axion may eventually be needed to solve the strong CP problem.

One important consequence of the intermediate scale  $M_{B-L}$ is
that the $\nu_i^c$ masses can be of order $10^8\ GeV$, which
facilitates leptogenesis to proceed in an inflationary
scenario$^{(7)}$. It also opens up the exciting
possibility that one of the `light' neutrinos has mass in the 3-10 eV
range, a feature hinted$^{(10)}$ at by the recent COBE data. However,
this would mean that the MSW$^{(11)}$ solution of the solar neutrino
problem cannot be implemented.

The inflationary scenario predicts that the density parameter
$\Omega$ is unity, and it has been suggested that $\Omega_\nu \approx
0.2 - 0.3$. That is, the bulk of the missing mass is in `cold' dark
matter. One of the major consequences of the $Z_2$ matter parity in
our model is the stability of the lightest supersymmetric particle
(LSP). The latter can be either the neutralino or possibly the
sneutrino.
In the
first case it has positive matter parity but it has to decay (by
fermion number conservation) to an odd number of negative matter
parity light fermions. In the second case it has negative matter
parity but has to decay to an even number of negative matter parity
light fermions. In both cases the decay process is forbidden by the
unbroken $Z_2$ matter parity. Therefore the LSP is a `cold' dark
matter candidate in our model.

In the second realization of the $SU(3)_c \times SU(3)_L \times
SU(3)_R$ model, we take $M_{B-L} = M_X$. Assuming that the $Q,
\bar{Q}$ and $ Q^c, \bar{Q}^c$ mass terms are $\frac{1} {3} M_X$ and the
$g, g^c $ 's acquire masses $\sim 10^{-1} M_X$
one loop renormalization group analysis yields
$(M_S \simeq 10^{3} GeV$)

\begin{equation}\begin{array}{rcl}
\sin^2\theta_W (M_Z) & \simeq & 0.23, \alpha_s(M_Z) \simeq 0.11,
\alpha_G(M_P) \simeq 0.20,\strut\\
M_X & = & M_{B-L} \simeq 10^{15} GeV.\end{array}
\end{equation}

\noindent
In this case, however, the pair of Higgs doublets $h^{(2)}, h^{(1)}$
does not necessarily belong to the same linear combination of
$\lambda_a$'s. Also, the light $e_i^c$'s and the `light' $\nu_i^c$'s
do not belong to the same three $\lambda_i$'s. This means that we
lose the asymptotic relations in eqs. (5) and (6). The `heavy'
neutrino masses are $M_{\nu_{i}^c} \sim M_{B-L}^2/M_P \sim 10^{11}
GeV$, and we  assume that they are all comparable. The `reheat' temperature
is again calculated from formulae (7) and (8), and turns out to
be much higher than $1 TeV$. However, taking into account the fact
that $M_{\nu^c}/<\phi>$ is an upper bound on the effective
$\phi \nu_i^c\nu_j^c$ coupling constants, the value of $T_r$ calculated
this way should be considered only as an upper bound.

The lepton asymmetry generated by the decaying $\nu_i^c$'s is given
by

\begin{equation}
\frac{n_L}{s} = \sum^3_{i = 1} \frac{n_{Li}}{s},
\end{equation}

\noindent
where the expression for $n_{Li}/s$ is given in eq. (9). All of the
$\nu_i^c$'s have comparable masses of order $10^{11}\ GeV$, and we can
assume that they are equally produced through the decay of the inflaton
field $\phi$ (with mass $m_\phi \stackrel{_>}{_\sim} 2\cdot 10^{11}\ GeV$).
Taking $(m_D)_{33}$ to be the largest element of $m_D$, the
lepton asymmetry, predominantly produced through the decays of
$\nu^c_{1,2}$, is estimated to be

\begin{equation}
\frac{n_L}{s} \sim \frac{1}{\pi} \frac{T_r}{m_\phi}
\left|\frac{(m_D)_{33}}{<h^{(1)}>}\right|^2 \; .
\end{equation}

\noindent
The allowed values for $(m_D)_{33}$
satisfy the following inequality

\begin{equation}
m_{\nu_{\tau}} \simeq \frac{|(m_D)_{33}|^2}{M_{\nu^{c}}}
\stackrel{_<}{_\sim} 10 eV \; .
\end{equation}

\noindent
Thus, eq.$(15)$ can easily lead to acceptable values for the baryon
asymmetry of the Universe.
The bound in eq.(16) is saturated for $(m_D)_{33} \simeq 30\ GeV$,
leading to a $\tau$-neutrino mass of about $10 eV$
which can provide the `hot' dark matter of the universe.
The LSP is again a `cold' dark matter candidate.

To conclude, we have outlined a possible alternative to minimal
supersymmetric $SU(5)$ unification. It offers the feature
that the proton is stable, and it accomplishes this in
a relatively elegant manner, avoiding in the process also some
unacceptable mass relations. Other advantages include the ease with
which the baryon asymmetry is generated (necessarily via
\nopagebreak[4] leptogenesis) and the presence of some hot dark matter in the
form of massive neutrinos.  Also, since the Higgs doublets and the
color triplet fields belong to different representations, the doublet
- triplet splitting can, in principle, be accomplished more easily than
in $SU(5)$.

We considered two alternative realizations of the model. In the first
case, the B-L breaking scale is much lower than the unification
scale. This leads to asymptotic relations like $m_t/m_b \simeq
\tan\beta$. The requirement of a sizable baryon asymmetry in this
case favors an unstable $\tau$-neutrino with mass of about $30\ MeV$ and
a lifetime of 20 sec.  The $\nu_\mu$-neutrino mass can be chosen so that it
provides either the `hot' dark matter of the universe or is
consistent with the MSW solution of the solar neutrino problem. In the second
realization of the model the B-L breaking scale coincides with the
unification scale. In this case a tau neutrino with mass of order $10 eV$
can provide some `hot' dark matter, while the muon neutrino plays a
role in the MSW effect. In either case a stable LSP provides for a `cold'
dark matter candidate.

\noindent
{\large\bf Acknowledgement:} We thank K.S. Babu for several important
discussions.

\section*{References}

\begin{enumerate}
\item S. Dimopoulos and H. Georgi, Nucl. Phys., \underline{B193}
(1981) 150;
N. Sakai, Z. Phys., \underline{C11} (1981) 153.

\item J. Ellis, S. Kelley and D.V. Nanopoulos, Phys. Lett.,
\underline{B249} (1990) 441;
V. Amaldi, W. de Boer and H. Furstenan, Phys. Lett., \underline{B260}
(1991) 447;
P. Langacker and M.X. Luo, Phys. Rev. D, \underline{D44} (1991) 817.

\item Kamiokande Collaboration (1992).

\item P. Candelas, G. Horowitz, A. Strominger and E. Witten, Nucl.
Phys., \underline{B258} (1985) 46;
E. Witten, Nucl. Phys., \underline{B258} (1985) 75.

\item Q. Shafi and X.M. Wang, 1990 (unpublished);
L. Ibanez and G.G. Ross, Nucl. Phys., \underline{B368} (1992) 3.

\item M. Fukugita and T. Yanagida, Phys. Lett., \underline{174B}
(1986) 45.

\item G. Lazarides and Q. Shafi, Phys. Lett., \underline{258B} (1991)
305.

\item V. Kuzmin, V. Rubakov and M. Shaposhnikov, Phys. Lett.,
\underline{B155} (1985) 36.

\item S.A. Bludman, Phys. Rev., \underline{D45} (1992) 4720.

\item R. Schaefer and Q. Shafi, Nature, \underline{359} (1992) 359.

\item L. Wolfenstein, Phys. Rev., \underline{D17} (1978) 2369; Phys.
Rev., \underline{D20} (1979) 2634; S.P. Mikheyev and A.Yu. Smirnov,
Usp. Fiz. Nauk., \underline{153} (1987) 3.

\end{enumerate}

\end{document}